\RequirePackage{fix-cm}
\documentclass[twocolumn,epjc3,preprint]{svjour3}
\smartqed  
\RequirePackage{graphicx}
\graphicspath{{./plots/}}
\newcommand{\lr}[1]{\left(#1\right)}

\newcommand{\lrs}[1]{\left[#1\right]}
\newcommand{\vev}[1]{\left\langle \, #1 \, \right\rangle}

\newcommand{\tr}[0]{\mathrm{Tr}}

\usepackage{amsmath}
\usepackage{slashed}
\usepackage[colorlinks=true]{hyperref}

\journalname{Eur. Phys. J. A}
\begin{document}
\sloppy

\title{Static magnetic susceptibility in finite-density $SU\lr{2}$ lattice gauge theory}
\titlerunning{Magnetic susceptibility in SU(2) gauge theory}        
\author{P.~V.~Buividovich \thanksref{e1,aLiverpool}
        \and
        D.~Smith\thanksref{aFair}
        \and
        L.~von~Smekal\thanksref{aGiessen,aHelmholz} 
}

\thankstext{e1}{e-mail:pavel.buividovich@liverpool.ac.uk}


\institute{Department of Mathematical Sciences, University of Liverpool, Liverpool, L69 7ZL, UK \label{aLiverpool}
           \and
           Institut f\"ur Theoretische Physik, Justus-Liebig-Universit\"at, 35392 Giessen, Germany \label{aGiessen}
           \and
           Facility for Antiproton and Ion Research in Europe GmbH (FAIR GmbH), 64291 Darmstadt, Germany \label{aFair}
           \and
           Helmholtz Research Academy Hesse for FAIR (HFHF), Campus Giessen, 35392 Giessen, German
           \label{aHelmholz}
}

\date{Received: date / Accepted: date}

\maketitle

\begin{abstract}
We study static magnetic susceptibility $\chi(T, \mu)$ in $SU(2)$ lattice gauge theory with $N_f = 2$ light flavours of dynamical fermions at finite chemical potential $\mu$. Using linear response theory we find that $SU(2)$ gauge theory exhibits paramagnetic behavior in both the high-temperature deconfined regime and the low-temperature confining regime. Paramagnetic response becomes stronger at higher temperatures and larger values of the chemical potential. For our range of temperatures $0.727 \leq T/T_c \leq 2.67$, the first coefficient of the expansion of $\chi\lr{T, \mu}$ in even powers of $\mu/T$ around $\mu=0$ is close to that of free quarks and lies in the range $(2 \ldots 5) \cdot 10^{-3}$. The strongest paramagnetic response is found in the diquark condensation phase at $\mu > m_{\pi}/2$.
\end{abstract}

\section{Introduction}
\label{sec:intro}

One of the fundamental quantities that characterize the response of some medium to the applied external magnetic field $\vec{H}$ is the \emph{magnetic susceptibility} $\chi$. It characterizes the magnetic field $\vec{H}_{int}$ created by spin polarization and electric currents that are induced in the medium by the external field $\vec{H}$. The magnetic field within the medium is $\vec{B} = \vec{H} + \vec{H}_{int} = \lr{1 + \chi} \vec{H}$, thus $\chi$ characterizes whether the external magnetic field is screened or enhanced within the medium. Medium with $\chi > 0$ is \emph{paramagnetic} and is attracted by magnetic field. Characteristic examples of paramagnetic media are metals like iron. Medium with $\chi < 0$ is \emph{diamagnetic} and is repelled by the magnetic field. Extreme examples of diamagnetic media are superconductors, for which $\chi = -1$ and hence the external magnetic field is completely screened.

Magnetic susceptibility of dense and hot QCD matter plays an important role in the dynamics of magnetar stars \cite{BlandfordMagneticSusceptibility,Endrodi:1407.1216}. Paramagnetism of QCD matter is also conjectured to lead to the magnetic ``squeezing'' of a fireball produced in off-central heavy-ion collisions \cite{Bali:1311.2559}, which should modify the observable elliptic flow upon hadronization.

The question of whether the QCD medium is paramagnetic or diamagnetic appears to be nontrivial, and the answer might depend on its temperature and density. At high temperatures, when quarks effectively behave as free Dirac fermions, the quark-gluon plasma is expected to be paramagnetic. This conclusion is also confirmed by lattice simulations in the high-temperature phase of QCD \cite{DElia:1307.8063,Levkova:1309.1142,Braguta:1909.09547}.

Calculations within the non-interacting hadron resonance gas model \cite{Endrodi:1301.1307} indicated that QCD matter is also paramagnetic in the hadronic phase below the deconfinement transition. However, first-principle lattice calculations also revealed signatures of weak diamagnetism at low temperatures \cite{Bali:1406.0269,Endrodi:1407.1216,Endrodi:2004.08778}. Diamagnetic response in the regime of weak magnetic fields was also found within the chiral perturbation theory \cite{Hofmann:2103.04937}. A change from diamagnetism for low-temperature QCD to paramagnetism at higher temperatures is predicted by the parton-hadron string dynamics (PHSD) model \cite{Cassing:1312.3189}, the functional renormalization group \cite{Kanazawa:1410.6253} and holographic QCD \cite{BallonBayona:2005.00500}.

In ongoing heavy-ion collision experiments both finite temperature and finite density play prominent roles. However, so far magnetic susceptibility of dense QCD matter has received somewhat less attention than its zero-density counterpart. A textbook knowledge is that for free fermions the magnetic susceptibility grows with density, and we can expect to observe a similar growth in QCD matter in the deconfined regime at sufficiently high temperatures, or at sufficiently large densities, where quarks are weakly interacting due to asymptotic freedom. A calculation within the hard thermal loop approximation confirms this expectation \cite{Ghosh:2103.08407}. Likewise, calculations within the holographic Sakai-Sugimoto model \cite{Bergman:0806.0366} (where pion-like degrees of freedom are present at low temperatures) suggest that magnetic susceptibility grows with density and therefore remains positive (paramagnetic) also at finite density. On the other hand, a zero-temperature, finite-density calculation within the Fermi liquid model \cite{Tatsumi:1008.3753} shows a nontrivial density dependence of magnetic susceptibility, with change of sign and singular behavior at some critical density. This behavior however appears to be quickly washed out due to thermal effects in favor of purely paramagnetic response.

An obvious obstacle for first-principle lattice studies of the magnetic susceptibility of QCD at finite chemical potential is the infamous fermionic sign problem. In this paper we study the effect of finite chemical potential on the magnetic susceptibility in $SU\lr{2}$ lattice gauge theory with $N_f = 2$ mass-degenerate light dynamical quarks, which is free of the fermionic sign problem at all values of the chemical potential \cite{Kogut:hep-lat/0105026,Kogut:hep-ph/0001171}. The Euclidean path integral of this gauge theory in the presence of an external electromagnetic field $A_{\mu}\lr{x}$ and a finite chemical potential $\mu$ reads
\begin{eqnarray}
\label{su2gt_path_integral}
 \mathcal{Z} = \int \mathcal{D}A_{\mu}^a \mathcal{D}\bar{\psi}_f \mathcal{D}\psi_f
 \nonumber \\
 \exp\lr{- \sum\limits_{f=u,d} \bar{\psi}_f \slashed{D}\lrs{A_{\mu}^a, A_{\mu}} \psi_f - S_{YM}\lrs{A_{\mu}^a}} ,
\end{eqnarray}
where $\psi_f$ are quark fields with flavour $f = u,d$ in the fundamental representation of $SU\lr{2}$ gauge group, $A_{\mu}^a$, $a = 1,2,3$ are $SU\lr{2}$ gauge fields in the adjoint representation of $SU\lr{2}$ and $S_{YM}\lrs{A_{\mu}^a}$ is the Yang-Mills action. External electromagnetic field and chemical potential enter the action via the Dirac operator
\begin{eqnarray}
\label{DiracOp}
 \slashed{D}\lrs{A_{\mu}^a, A_{\mu}} = \gamma_{\mu} \lr{\partial_{\mu} - i A_{\mu}^a \sigma_a - i A_{\mu}} + m + \mu \, \gamma_0 ,
\end{eqnarray}
where $\gamma_{\mu}$ are the Dirac $\gamma$-matrices, $\sigma_a$ are the Pauli matrices in $SU\lr{2}$ color space, $m$ is the bare quark mass (assumed to be the same for both quark flavors), and $\mu$ is the chemical potential.

$SU\lr{2}$ gauge theory is expected to be qualitatively similar to QCD at sufficiently small values of chemical potential $\mu < m_{\pi}/2$. Similarly to QCD, in this regime $SU\lr{2}$ theory undergoes a crossover between the low-temperature confining regime with spontaneously broken chiral symmetry and the high-temperature deconfinement regime with restored chiral symmetry. On the other hand, a theory with $SU\lr{2}$ gauge group and $N_f = 2$ quark flavors has five distinct pion states, in contrast to three pions in real QCD \cite{Kogut:hep-ph/0001171}. As the chemical potential becomes larger than half of the pion mass, $\mu > m_{\pi}/2$, $SU\lr{2}$ gauge theory enters the diquark condensation phase which is absent in real QCD. Therefore qualitative similarity to QCD is lost at $\mu > m_{\pi}/2$. However, very deep in the diquark condensation phase and at low temperatures, the physics of $SU\lr{2}$ gauge theory is expected to resemble that of the conjectured quarkyonic phase \cite{Pisarski:0706.2191,Braguta:1605.04090}.

Another conceptually similar approach to avoid the sign problem is to study $SU\lr{3}$ gauge theory, but at finite isospin chemical potential $\mu_I$ \cite{Son:hep-ph/0005225}. The dependence of magnetic susceptibility on finite isospin chemical potential was studied in first-principle lattice simulations in \cite{Endrodi:1407.1216}. It was found that pion condensation leads to relatively strong diamagnetic response at $\mu_I > m_{\pi}/2$ and low temperatures. It is expected that at very large $\mu_I$ and/or sufficiently high temperatures the paramagnetic behavior should reappear again due to asymptotic freedom \cite{Endrodi:1407.1216}.

In agreement with previous studies for $SU\lr{3}$ gauge theory \cite{DElia:1307.8063,Levkova:1309.1142,Braguta:1909.09547,Endrodi:2004.08778}, in our study we find that $SU\lr{2}$ gauge theory is paramagnetic in the high-temperature regime. We also find a weak paramagnetic response in the low-temperature confining regime. At all temperatures the finite chemical potential appears to make the paramagnetic response stronger.

\section{Numerical measurements of magnetic susceptibility within the linear response approximation}
\label{sec:method}

The QCD magnetic susceptibility is often calculated in terms of the response of a free energy to an external magnetic field, which is quantized in a finite volume \cite{DElia:1307.8063,Levkova:1309.1142,Braguta:1909.09547}. However, finite magnetic field breaks time-reversal invariance and therefore leads to the appearance of the fermionic sign problem even for finite-density $SU\lr{2}$ gauge theory. We therefore base our measurements on gauge field configurations generated without external magnetic field, and use linear response theory with respect to slowly varying weak magnetic field (with zero total flux across the lattice) to find the magnetic susceptibility. For similar reason, same approach was used also in the lattice study of magnetic susceptibility at finite isospin density \cite{Endrodi:1407.1216}.

Within the linear response theory, magnetic susceptibility is related to static transverse correlator of space-like electric currents in Euclidean (imaginary time) space \cite{GiulianiVignaleElectronLiquid,Endrodi:1407.1216,Endrodi:2004.08778}. For isotropic space this correlator can be written in the spatial momentum space as
\begin{eqnarray}
\label{jj1}
 \Pi_{kl}\lr{q} = \int d x_0 \int d^3 \vec{x} \, \vev{j_k\lr{x_0, \vec{x}} j_l\lr{0, \vec{0}}} e^{i \vec{q} \cdot \vec{x}}
 = \nonumber \\ =
 \lr{q^2 \delta_{kl} - q_k \, q_l} \, \Pi\lr{q^2} ,
\end{eqnarray}
where the electric current includes the contributions from all $N_f$ quark fields $\psi_f$ with appropriate charge factors $q_{u} = +2/3$, $q_{d} = -1/3$ for each flavor $f$:
\begin{eqnarray}
\label{jdef}
 j_k\lr{x} = \sum\limits_{f=1}^{N_f} q_f \, \bar{\psi}_f \gamma_k \psi_f .
\end{eqnarray}
To represent the raw lattice data, we also consider the spatial current-current correlators that are summed over all coordinates except for one of the spatial coordinates, say, $x_3$:
\begin{eqnarray}
\label{spatial_corr}
 G_{11}\lr{x_3} = \int \frac{d q_3}{2 \pi} e^{-i q_3 x_3} \Pi_{11}\lr{q_3^2}
 = \nonumber \\ =
  \int dx_0 dx_1 dx_2 \vev{j_1\lr{x_0, \vec{x}} j_1\lr{0, \vec{0}}} .
\end{eqnarray}

The magnetic susceptibility with respect to static magnetic fields in the long-wavelength limit is defined as \cite{GiulianiVignaleElectronLiquid,Endrodi:2004.08778}
\begin{eqnarray}
\label{chi_def}
 \chi_0 = \lim\limits_{q \rightarrow 0} \Pi\lr{q^2} .
\end{eqnarray}
To extract $\Pi\lr{q^2}$ from current-current correlators (\ref{jj1}) we take the momentum $\vec{q} = \lr{0, 0, q_3}$ in the direction of $x_3$ coordinate axis, and consider the momentum-space correlator $\Pi_{11}\lr{q_3} =  q_3^2 \Pi\lr{q_3^2}$. In this case we can find the magnetic susceptibility as
\begin{eqnarray}
\label{susc_def_bare}
  \chi_0 = \lim\limits_{q_3 \rightarrow 0} q_3^{-2} \, \Pi_{11}\lr{q_3^2} .
\end{eqnarray}

The susceptibility $\chi_0$ in (\ref{susc_def}) is the bare susceptibility that has to be renormalized to ensure that the magnetic susceptibility of QCD vacuum at zero temperature and density has its physical zero value. To this end one subtracts the value of $\chi_0\lr{T=0,\mu=0}$ from $\chi_0\lr{T, \mu}$ to obtain the physical susceptibility $\chi\lr{T, \mu}$ at temperature $T$ and chemical potential $\mu$:
\begin{eqnarray}
\label{susc_def}
 \chi\lr{T, \mu} = \chi_0\lr{T, \mu} - \chi_0\lr{T=0, \mu=0}.
\end{eqnarray}
In practice, lattice QCD simulations cannot reach zero temperature, and we subtract the value of $\chi_0 $ at the lowest temperature $T = \frac{1}{L_t \, a}$ with $L_t = 22$ used in our simulations.

The most important contribution to the current-current correlator (\ref{jj1}) comes from connected fermionic diagrams. For conserved electromagnetic currents on the lattice, this contribution can be represented as
\begin{eqnarray}
\label{current_current_connected}
 \vev{j_{x,k} j_{y,l}}_{conn}
 = \nonumber \\ =
 \sum\limits_f q_f^2 \, \left. \tr\lr{
  \frac{\partial D}{\partial \theta_{x,k}} \frac{1}{D}
  \frac{\partial D}{\partial \theta_{y,l}} \frac{1}{D}
 } \right|_{\theta=0}
 + \nonumber \\ +
 \sum\limits_f q_f^2 \, \delta_{xy}\delta_{kl}
 \left. \tr\lr{\frac{\partial^2 D}{\partial \theta_{x,k}^2} D^{-1}}\right|_{\theta=0} ,
\end{eqnarray}
where $x$, $y$ are now the sites of the four-dimensional lattice, and $D$ is the Dirac operator with both non-Abelian gauge fields and an $U\lr{1}$ lattice field $\theta_{x,\mu}$, with link factors $e^{i \theta_{x,\mu}}$. Since we work with conserved lattice currents, our results for magnetic susceptibility should not be renormalized, apart from the subtraction of the vacuum susceptibility in (\ref{susc_def}).

The last term in (\ref{current_current_connected}) is the contact term that just adds a $q$-independent constant to the current-current correlator (\ref{jj1}). We have found that this constant exactly cancels the finite value of the Fourier transform of the first summand in (\ref{current_current_connected}) at $q = 0$ for each gauge field configuration, so that the limit $\lim\limits_{q \rightarrow 0} q^{-2} \, \Pi_{11}\lr{q}$ in (\ref{susc_def_bare}) becomes well-defined. This cancellation is not accidental and ensures the finiteness of the magnetic susceptibility in the long-wavelength limit. For this reason we do not measure the contact term in our simulations. Instead, we measure only the first summand in (\ref{current_current_connected}) and obtain its Fourier transform $\bar{\Pi}_{11}\lr{q_3^2}$. The corresponding space-averaged current-current correlator, obtained by replacing $\Pi_{11}\lr{q_3^2}$ with $\bar{\Pi}_{11}\lr{q_3^2}$ in (\ref{spatial_corr}) is denoted as $\bar{G}_{11}\lr{x_3}$. $\bar{\Pi}_{11}\lr{q_3^2}$ is finite at $q_3 = 0$ and is symmetric around this point. It can therefore be expanded around $q_3 = 0$ as
\begin{eqnarray}
\label{Pi11_exp1}
 \bar{\Pi}_{11}\lr{q_3^2} = A - q_3^2 \, B + O\lr{q_3^4} .
\end{eqnarray}

The constant $A$ is cancelled by the contribution of the contact term, and inserting the above decomposition into (\ref{susc_def}) we conclude that the bare magnetic susceptibility $\chi_0$ is given by $-B$, and we can rewrite (\ref{Pi11_exp1}) as
\begin{eqnarray}
\label{chi_numdef}
 \bar{\Pi}_{11}\lr{q_3^2} = A + \chi_0 \, q_3^2 + O\lr{q_3^4} .
\end{eqnarray}
We can therefore also express the magnetic susceptibility $\chi_0$ in terms of the second derivative of $\bar{\Pi}_{11}\lr{q_3^2}$ with respect to $q_3$:
\begin{eqnarray}
\label{chi_numdef_derivative}
 \chi_0 = \frac{1}{2} \frac{d^2}{d q_3^2} \bar{\Pi}_{11}\lr{q_3^2} .
\end{eqnarray}
In practice we construct the interpolating polynomial using the discrete values of $\bar{\Pi}_{11}\lr{q_3^2}$ at five lowest momenta $q_3 = \frac{2 \pi \, k}{L_s}$ for $k = \pm 2, \, \pm 1, 0$, and find the second derivative in (\ref{chi_numdef_derivative}) as the second derivative of this interpolating polynomial. Statistical errors of $\chi_0$ are estimated using bootstrapping.

The current-current correlators in (\ref{jj1}) also contain the contribution of disconnected fermionic diagrams, see Section IV of \cite{Buividovich:20:1} for an explicit expression. This contribution is however typically very small and difficult to measure. For this reason with only consider the connected contribution (\ref{current_current_connected}) in this work. Let us also note that if the diquark source $\lambda$ is nonzero, the expressions for current-current correlators become somewhat more complicated than (\ref{current_current_connected}). We again refer the reader to Appendix~C of \cite{Buividovich:20:1} for explicit expressions.

\section{Lattice setup}
\label{sec:lattice_setup}

For the measurements reported in this work we use the same set of lattice configurations with spatial lattice size $L_s = 30$ that was used in our recent papers \cite{Buividovich:20:1,Buividovich:20:2}. To make the paper self-contained, let us briefly summarize here the most important details of our lattice action.

We use the standard Hybrid Monte-Carlo algorithm with $N_f=2$ mass-degenerate rooted staggered fermions and a tree-level improved Symanzik gauge action to generate gauge field configurations. The bare mass of staggered fermions is $a m_{stag} = 0.005$, which corresponds to the pion mass $a m_{\pi} = 0.158 \pm 0.002$ and the ratio of pion and $\rho$-meson masses $m_{\pi}/m_{\rho} \approx 0.4$. We work at a fixed gauge coupling $\beta = 1.7$, hence at fixed lattice spacing.

To improve momentum resolution in the measurements of $\bar{\Pi}_{11}\lr{q_3^2}$, we use lattices with spatial size $L_s = 30$. Temperature is varied by changing the temporal lattice size $L_t$ between $L_t = 6$ and $L_t = 22$ in steps of two. We consider three distinct values of the chemical potential $a \mu = 0.0, \, 0.05, \, 0.20$, of which the first two are below the pion condensation threshold. For $L_t > 12$ we generate gauge configurations with a small diquark source $a \lambda = 5 \cdot 10^{-4}$ that serves as a seed for diquark condensation in a finite volume. To measure the magnetic susceptibility, at each value of $T$ and $\mu$ we use between 600 (high $T$, small $\mu$) and 100 (low $T$, large $\mu$) lattice configurations.

The phase diagram of $SU\lr{2}$ gauge theory within this lattice setup was studied in detail in \cite{Buividovich:20:1}. Ensembles of gauge configurations with $a \mu = 0.0$ and $a \mu = 0.05$ are in the QCD-like regime, in which the crossover towards the phase with spontaneously broken chiral symmetry occurs around $L_t = 16$. Ensembles with $a \mu = 0.2$ already have $\mu > m_{\pi}/2$, and are in the diquark condensation phase at sufficiently low temperatures with $L_t < 20$.

\begin{figure}[h!tpb]
  \centering
  \includegraphics[angle=-90,width=0.48\textwidth]{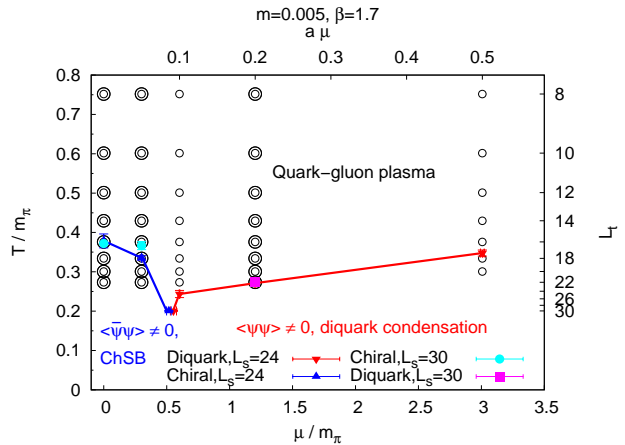}
  \caption{Phase diagram of finite-density $SU\lr{2}$ lattice gauge theory with $N_f = 2$ light quark flavors (plot taken from \cite{Buividovich:20:1}). Ensembles of gauge field configurations used in this work are shown as double empty circles.}
  \label{fig:phase_diagram}
\end{figure}

Current-current correlators in (\ref{jj1}) are measured using the Wilson-Dirac valence quarks with HYP-smeared gauge fields \cite{Hasenfratz:hep-lat/0103029}. Bare mass $m_{WD}=-0.21$ in the Wilson-Dirac operator is tuned in such a way that the pion mass measured with Wilson-Dirac quarks coincides with the pion mass for staggered quarks. While it is certainly also possible to calculate current-current correlators for staggered quarks, in this work we re-use the Wilson-Dirac current-current correlators used in our papers \cite{Buividovich:20:1,Buividovich:20:2}. The use of Wilson-Dirac valence quarks in these papers was motivated by the need to have a reasonably good definition of axial current in addition to the vector current.

\begin{figure*}[h!tpb]
 \includegraphics[width=0.48\textwidth]{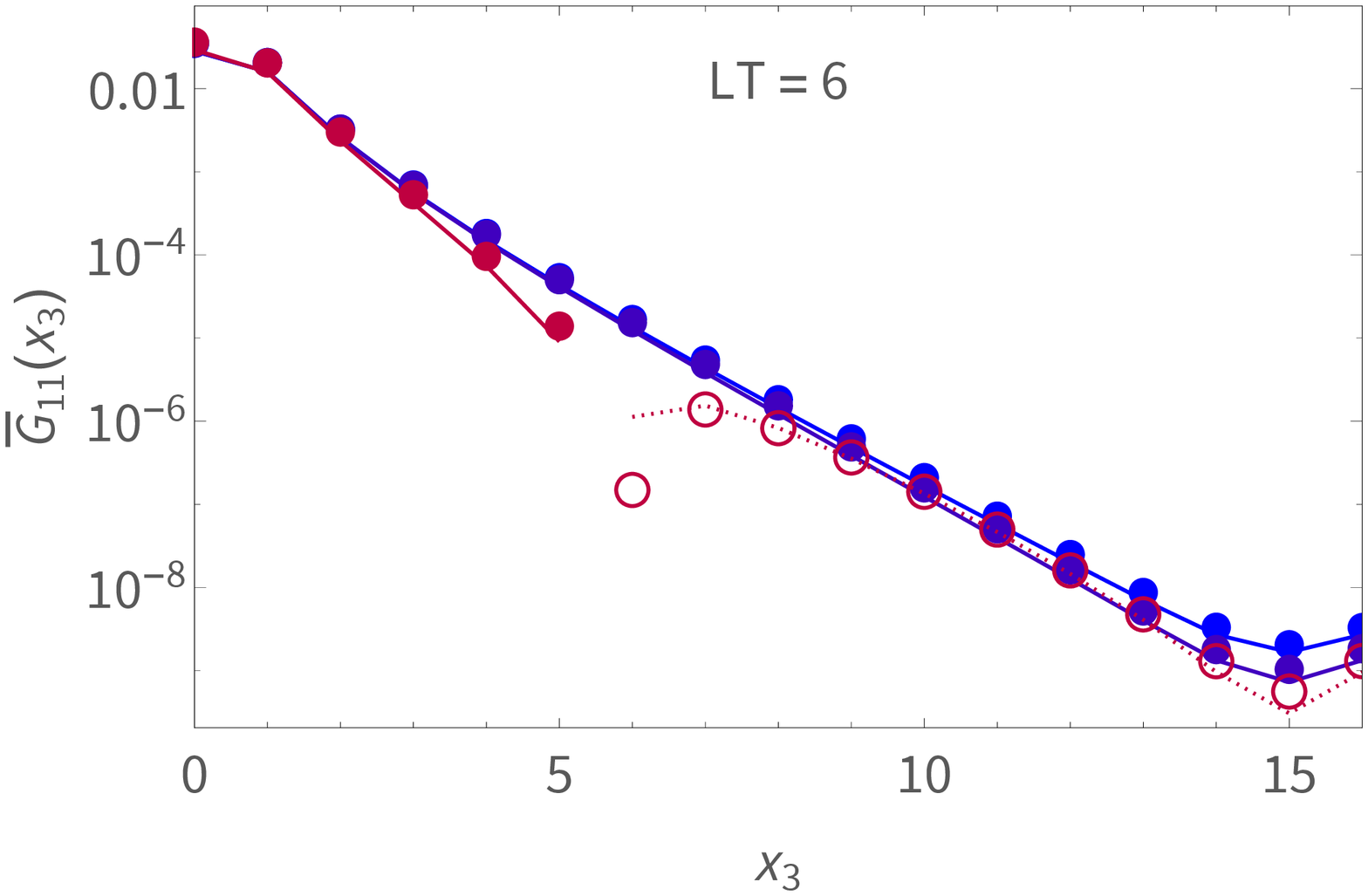}
 \includegraphics[width=0.48\textwidth]{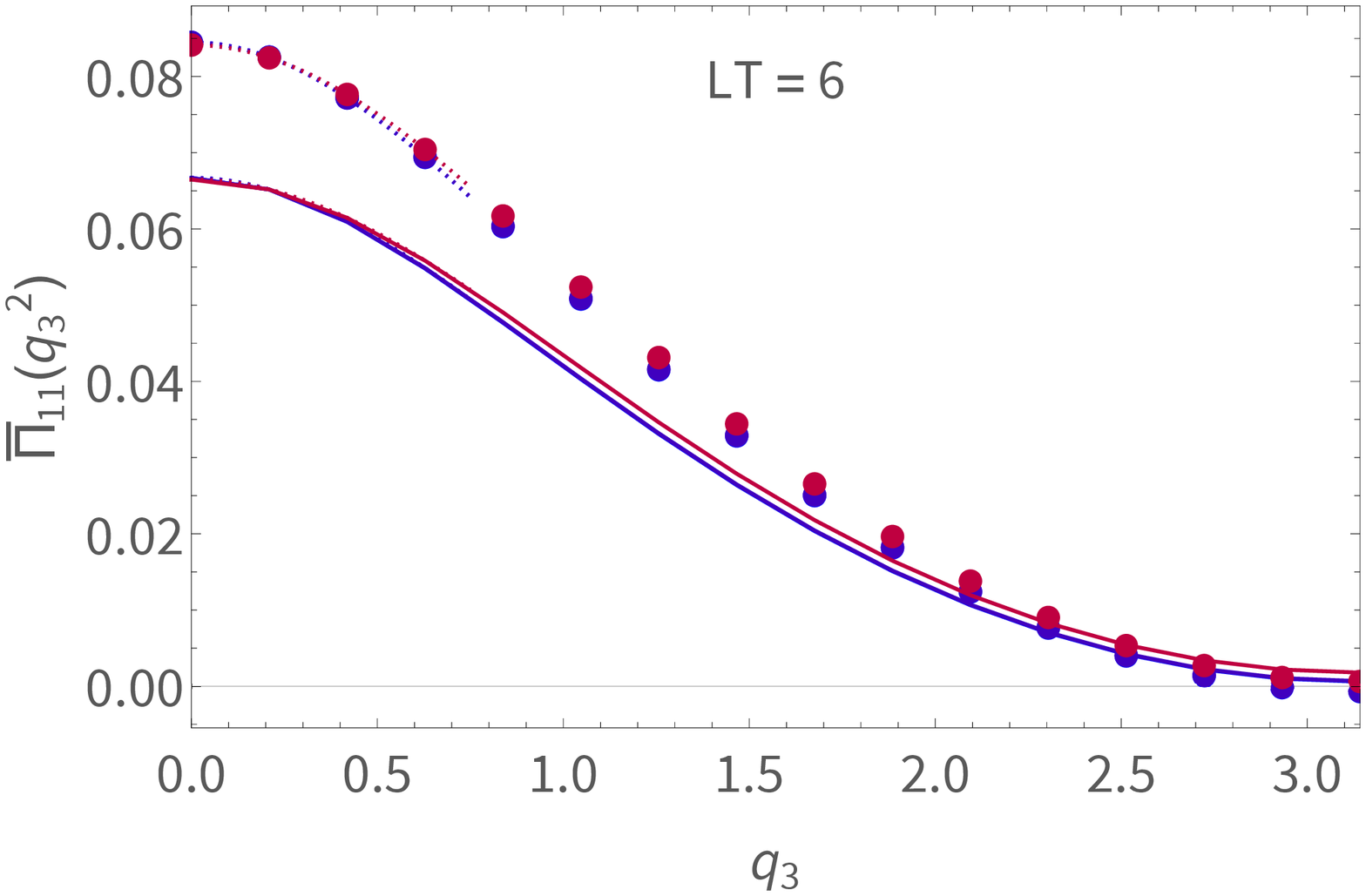}\\
 \includegraphics[width=0.48\textwidth]{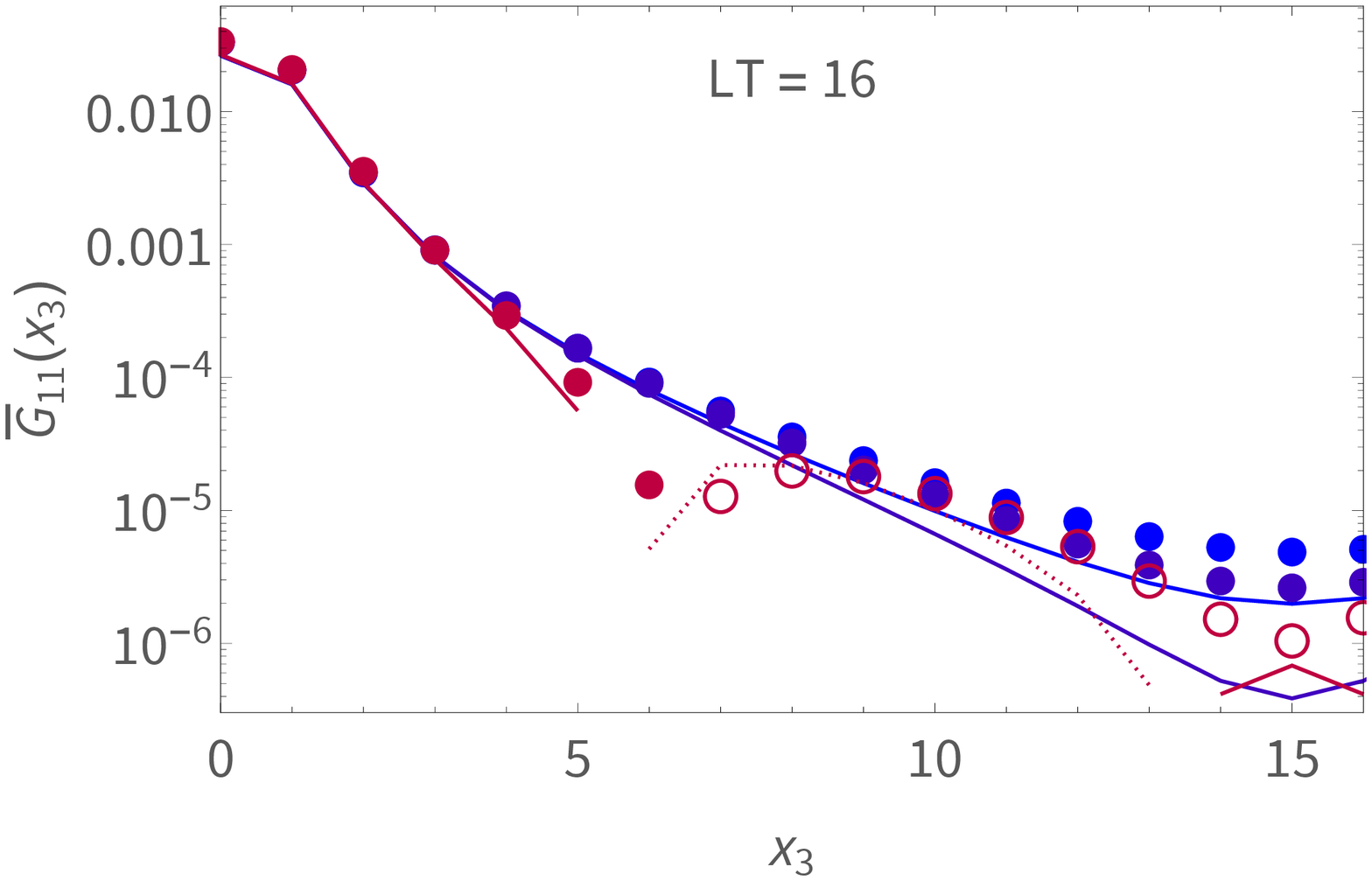}
 \includegraphics[width=0.48\textwidth]{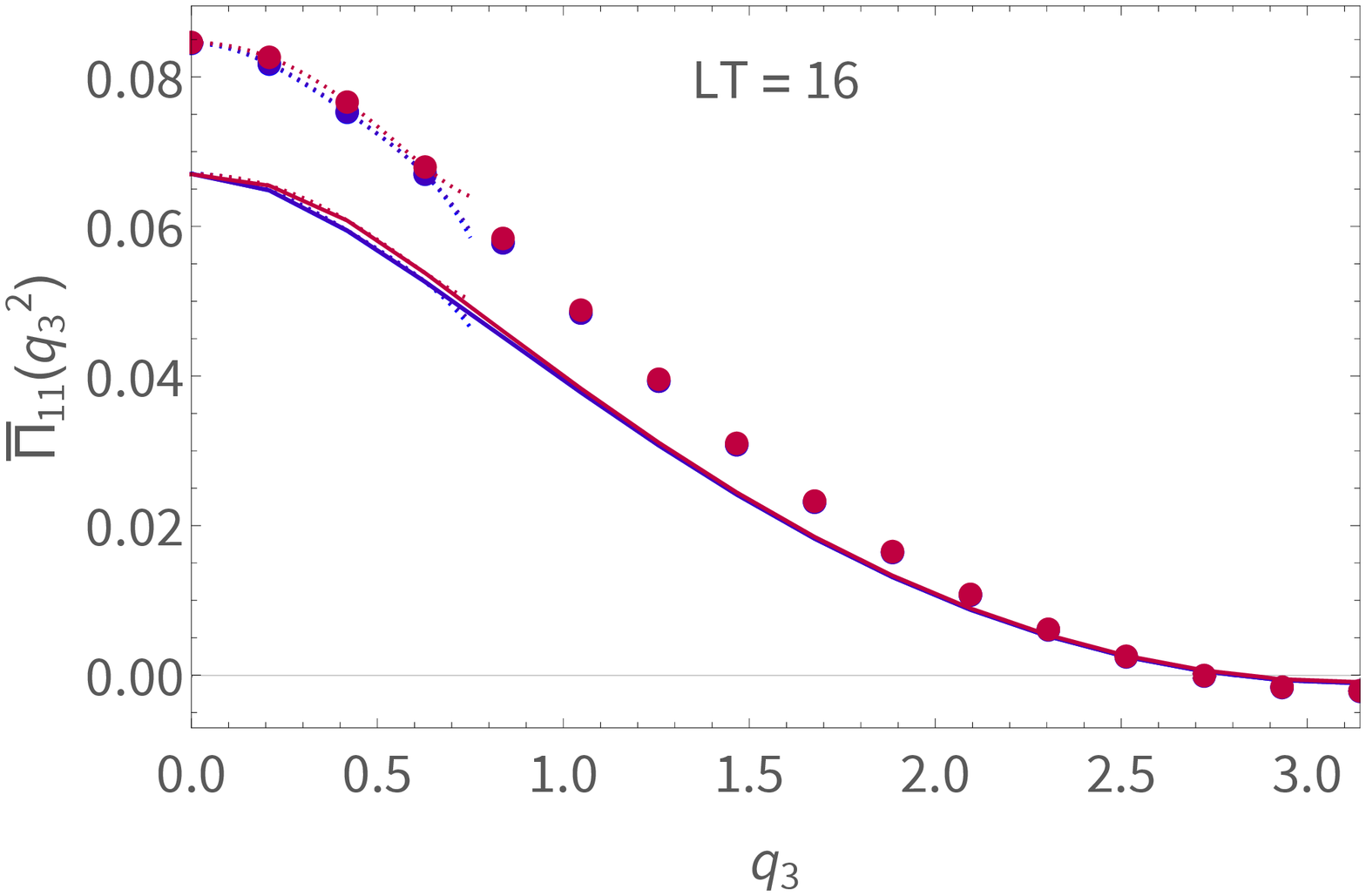}\\
 \includegraphics[width=0.48\textwidth]{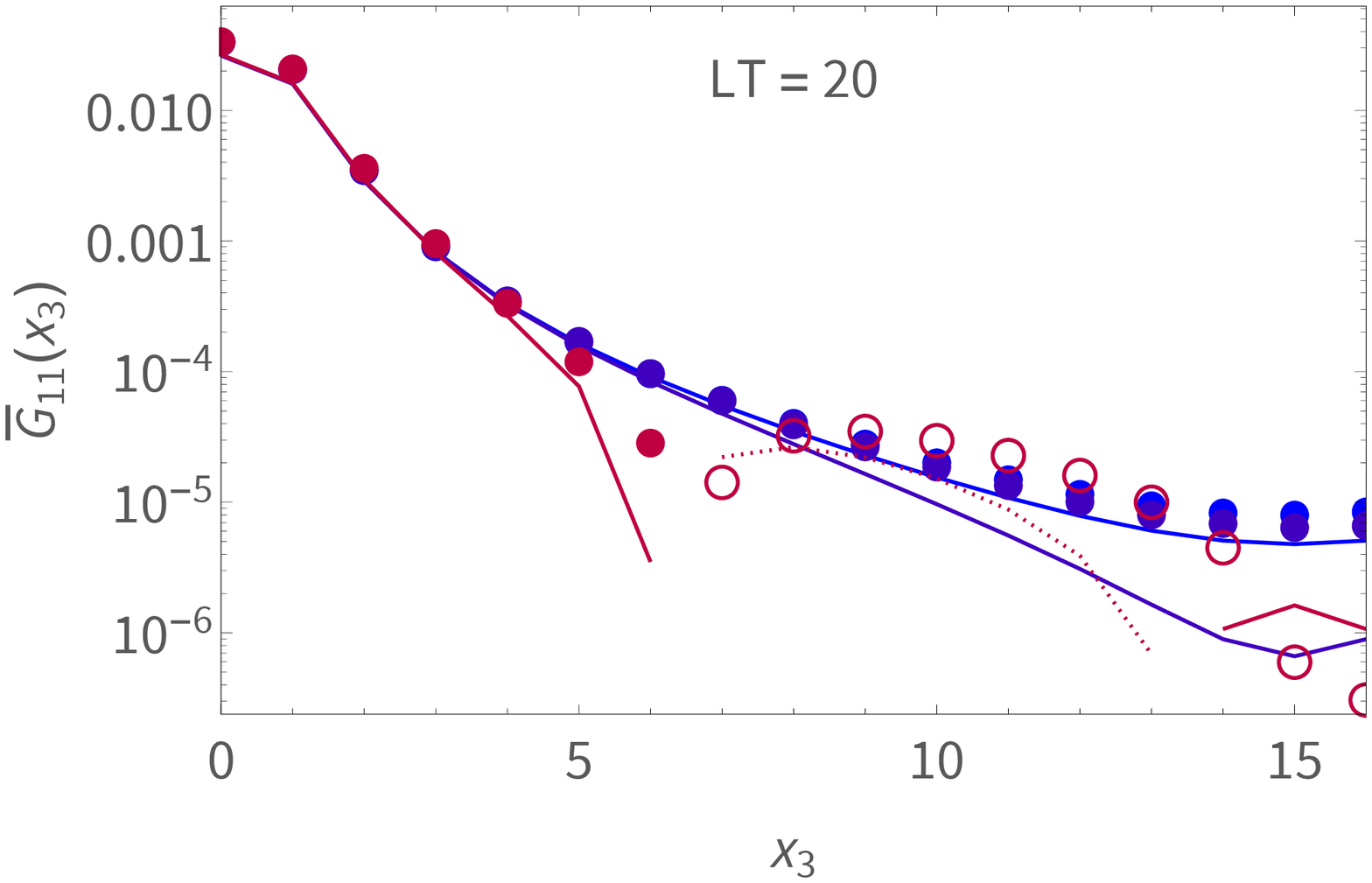}
 \includegraphics[width=0.48\textwidth]{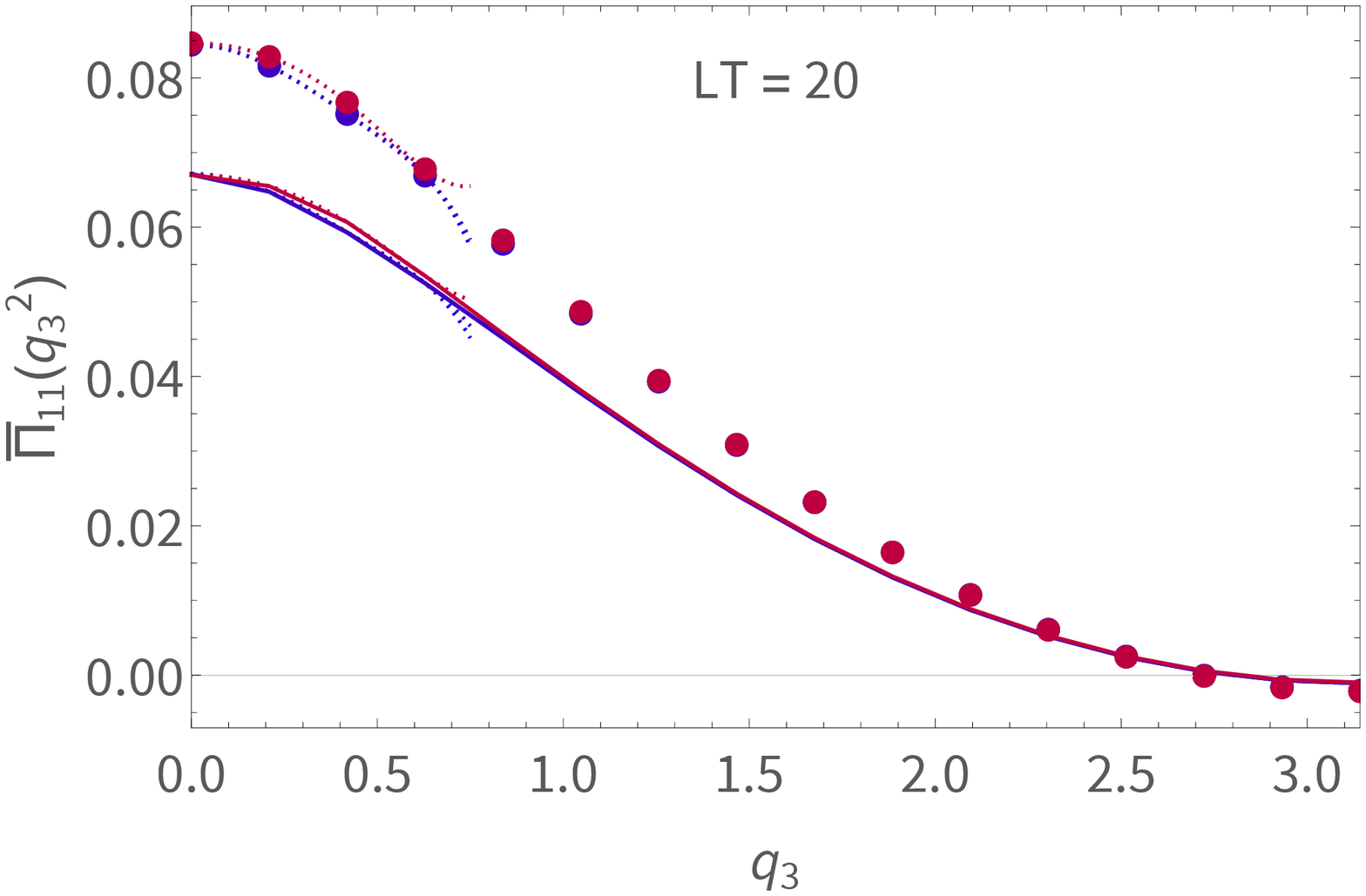}\\
\caption{Static correlators $\bar{G}_{11}\lr{x_3}$ and $\bar{\Pi}_{11}\lr{q_3^2}$ of spatial vector currents as functions of the spatial coordinate $x_3$ (Eq.~(\ref{spatial_corr})) and the spatial momentum $q_3$ (Eq.~(\ref{jj1})) at three different temporal lattice sizes: $L_0 = 6$ ($T > T_c$), $L_0 = 16$ ($T \approx T_c$) and $L_0 = 22$ ($T < T_c$). Solid lines show the free-fermion results obtained with the same lattice setup. All quantities are given in units of lattice spacing $a$. Interpolating polynomials used to obtain the magnetic susceptibility from the relation (\ref{chi_numdef_derivative}) are shown with dotted lines on the plots of $\bar{\Pi}_{11}\lr{q_3^2}$. For plots of $\bar{G}_{11}\lr{x_3}$, dotted lines represent the absolute value of $\bar{G}_{11}\lr{x_3}$ if $\bar{G}_{11}\lr{x_3} < 0$.}
\label{fig:jvjv}
\end{figure*}

\section{Numerical results}
\label{sec:numres}

In the left column on Fig.~\ref{fig:jvjv} we present the raw lattice data for the space-averaged current-current correlators $\bar{G}_{11}\lr{x_3}$. At high temperatures we observe a characteristic exponential decay of $\bar{G}_{11}\lr{x_3}$, which becomes somewhat less pronounced at lower temperatures. Deviations from free fermion results (shown with solid lines) become clearly larger towards lower temperatures. Interestingly, at $a \mu = 0.2$ both free fermion correlators and gauge theory correlators become negative at large $x_3$. On our logarithmic-scale plot on Fig.~\ref{fig:jvjv} we show the absolute value of these negative correlators using empty symbols (for lattice gauge theory data) and dashed lines (for free fermion results).

\begin{figure}[h!tpb]
 \includegraphics[width=0.48\textwidth]{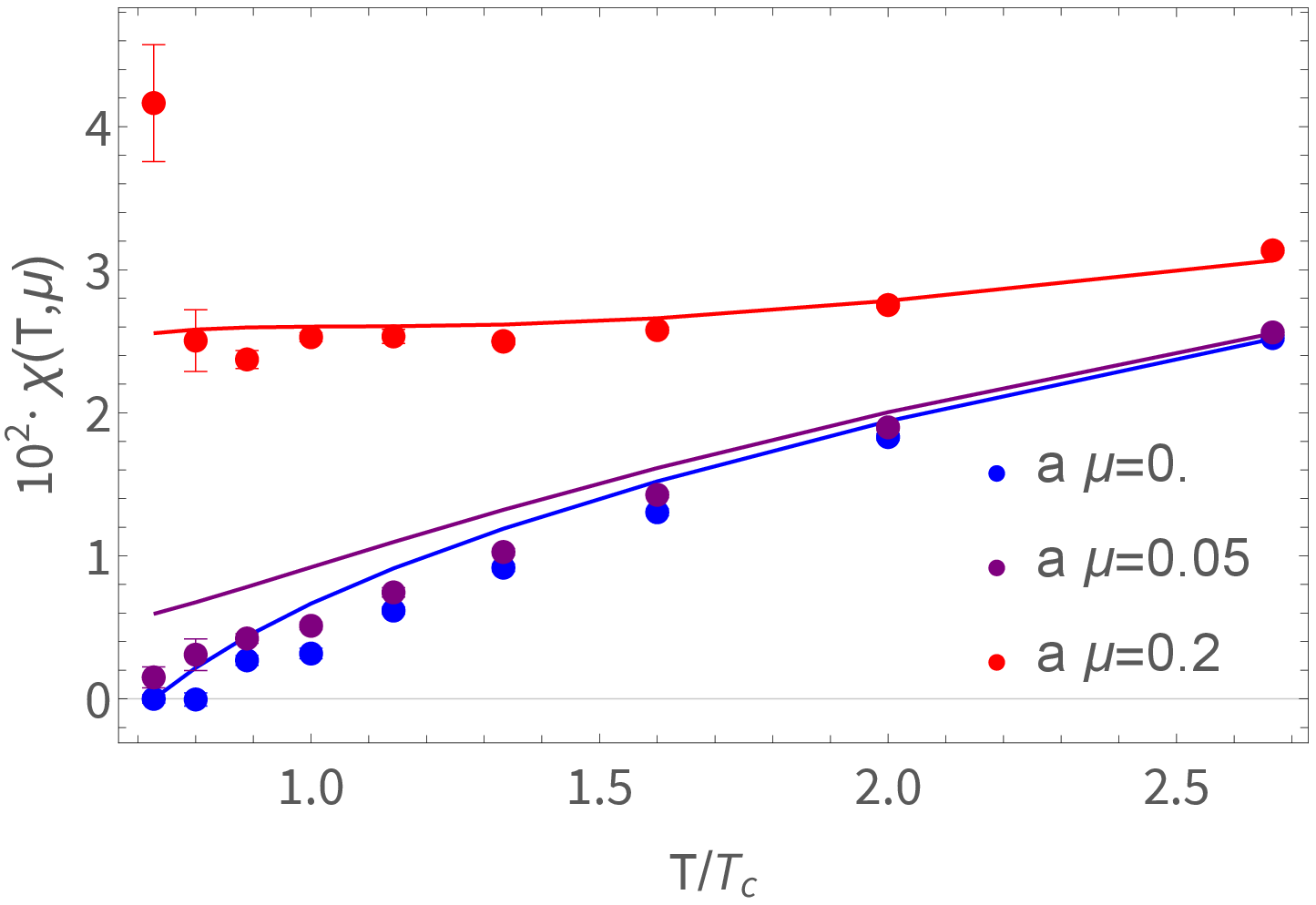}\\
\caption{Magnetic susceptibility $\chi$ as a function of temperature at the chemical potential $a \, \mu=0.0$, $a \, \mu = 0.05$ and $a \, \mu = 0.2$. The diquark condensation threshold is $a \, \mu = a m_{\pi}/2 = 0.079$. Solid lines show the free quark results for the same lattice parameters.}
\label{fig:chi}
\end{figure}

Fourier transforms $\bar{\Pi}_{11}\lr{q_3^2}$ of $\bar{G}_{11}\lr{x_3}$ all have bell-shaped form, with apparently small differences between the results at different temperatures and chemical potentials. However, these small differences become very essential once we subtract the contact term contribution (last term in (\ref{current_current_connected})) and the vacuum value of the bare susceptibility $\chi_0$ (see equation (\ref{susc_def})).

As discussed in Section~\ref{sec:method}, we construct an interpolating polynomial for the correlator $\bar{\Pi}_{11}\lr{q_3^2}$ using five data points that correspond to the smallest lattice momenta and calculate the bare magnetic susceptibility as half the second derivative of this polynomial, see (\ref{chi_numdef_derivative}). We obtain the renormalized magnetic susceptibility $\chi\lr{T, \mu}$ by subtracting the value of $\chi_0 = -0.07050 \pm 0.00027$ at $\mu = 0$ and $T = \frac{1}{22 \, a}$ ($L_t = 22$), the lowest temperature that we have.

While this temperature is not very small in comparison with the deconfinement temperature $T_c \approx \frac{1}{16 \, a}$, previous lattice simulations \cite{Endrodi:2004.08778} indicate very weak temperature dependence of $\chi_0$ in the low-temperature regime. To check this independently, we have also measured $\chi_0$ at a very low temperature $T = \frac{1}{56 \, a}$ on the $28^3 \times 56$ lattice, obtaining $\chi_0 = -0.0702 \pm 0.0007$. This result coincides with the one on $30^3 \times 22$ lattice within statistical errors. We still use $\chi_0$ calculated on $30^3 \times 22$ lattice for subtraction, because it has smaller statistical uncertainty and refers to the same lattice size as other data points

The resulting dependence of the magnetic susceptibility $\chi\lr{T, \mu}$ on temperature and chemical potential is illustrated on Fig.~\ref{fig:chi}. We plot the susceptibility as a function of the ratio $T/T_c$, where $T_c \approx \frac{1}{16 a}$ is the crossover temperature in our lattice setup.

We observe that below the diquark condensation threshold, at $\mu < m_{\pi}/2$, the magnetic susceptibility is positive and monotonically grows with temperature both at $T < T_c$ and $T > T_c$, approaching the magnetic susceptibility of free quarks at high temperatures. We observe no direct signatures of weak diamagnetism at low temperatures. However, for the second-lowest temperature $T = \frac{1}{20 \, a}$ ($L_t = 20$) $\chi\lr{T, \mu}$ appears to be zero within error bars. It was stressed in \cite{Endrodi:2004.08778} that extrapolation to the continuum limit $a \rightarrow 0$ is essential to observe the diamagnetic behavior at low temperatures. Since we work in the fixed-scale approach, we cannot exclude that once the data is extrapolated to $a \rightarrow 0$, $SU\lr{2}$ gauge theory might also exhibit a weak diamagnetic response. On the other hand, in the paramagnetic regime the susceptibility tends to slightly decrease towards the continuum limit \cite{Endrodi:1407.1216,Endrodi:2004.08778,Nikolaev:2008.12326}. We therefore expect that our result at finite lattice spacing might be slightly larger than the corresponding continuum limit.

\begin{figure}
  \centering
  \includegraphics[width=0.48\textwidth]{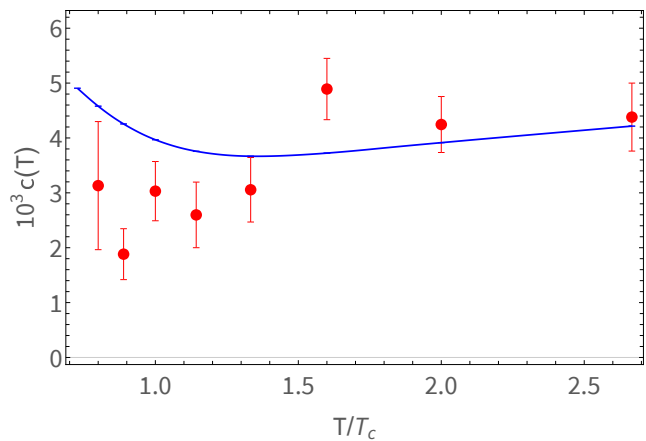}
  \caption{First nontrivial coefficient in the expansion of magnetic susceptibility $\chi\lr{T, \mu}$ in even powers of $\mu/T$ around $\mu/T = 0$, calculated from the difference of the data at $\mu = 0$ and $a \mu = 0.05$ according to (\ref{chi_expansion_findif}). Solid line shows the corresponding free fermion result calculated in the same way for the same lattice setup.}
  \label{fig:chi_expansion}
\end{figure}

Overall, our results for the magnetic susceptibility at zero density and sufficiently high temperatures are in good agreement with lattice QCD results \cite{Bali:1406.0269,Endrodi:2004.08778}, and are noticeably larger than the estimates obtained within the PHSD model \cite{Cassing:1312.3189}.

Small but finite chemical potential $\mu < m_{\pi}/2$ appears to increase the magnetic susceptibility at all temperatures and thus make the paramagnetic response stronger. At low temperatures, the dependence of $\chi\lr{T, \mu}$ on $\mu$ is weaker than for free quarks.

At small values of $\mu$ we can also expand $\chi\lr{T, \mu}$ in powers of $\mu/T$ around $\mu/T = 0$:
\begin{eqnarray}
\label{chi_expansion}
 \chi\lr{T, \mu} = \chi\lr{T, 0} + c_{\chi}\lr{T} \lr{\frac{\mu}{T}}^2 .
\end{eqnarray}
Due the charge conjugation symmetry of $SU\lr{2}$ gauge theory, this expansion only contains even powers of $\mu$. We estimate the coefficient $c_{\chi}\lr{T}$ from the data points at $\mu = 0$ and at our lowest nonzero value $\mu_1 = 0.05 \, a^{-1}$ as
\begin{eqnarray}
\label{chi_expansion_findif}
 c_{\chi}\lr{T} \approx \frac{T^2}{\mu_1^2} \lr{\chi\lr{T, \mu_1} - \chi\lr{T, 0}} .
\end{eqnarray}
We show the temperature dependence of $c_{\chi}\lr{T}$ on Fig.~\ref{fig:chi_expansion} together with the corresponding free fermion result.

For large values of the chemical potential $\mu > m_{\pi}/2$, the paramagnetic response becomes particularly strong. Interestingly, in this regime $\chi\lr{T, \mu}$ has rather weak temperature dependence, except for the data point at lowest temperatures. As one can see from Fig.~\ref{fig:phase_diagram}, this data point is in the diquark condensation phase. This observation suggests that diquark condensation phase is strongly paramagnetic.

Since in $SU\lr{2}$ gauge theory the quark chemical potential and the isospin chemical potential are equivalent \cite{Kogut:hep-ph/0001171}, it is instructive to compare our results with the lattice study \cite{Endrodi:1407.1216} of magnetic susceptibility at finite isospin chemical potential $\mu_I$ and at low temperatures. This study found that as $\mu_I$ reaches the pion condensation threshold, diamagnetic response becomes much stronger than at $\mu_I \ll m_{\pi}/2$, as can be expected for charged scalar bosons. However, at even larger $\mu_I$ one approaches the asymptotic freedom regime and the paramagnetic response was conjectured to set in again \cite{Endrodi:1407.1216}. We can also expect that diquark condensate, being a condensate of charged bosons, will exhibit a diamagnetic response at least at low temperatures and for some range of $\mu$ values with $\mu > m_{\pi}/2$. While we observe only a paramagnetic response, we cannot exclude diamagnetism immediately above the threshold. It might be that with our only value of chemical potential $a \, \mu = 0.2$ exceeding $a \, m_{\pi}/2 = 0.079$ we are already missing the diamagnetic regime.

\section{Conclusions}
\label{sec:conclusions}

We have used linear response theory to study the magnetic susceptibility $\chi\lr{T, \mu}$ of $SU\lr{2}$ gauge theory with $N_f = 2$ light quark flavours at finite temperature and density. In agreement with lattice QCD results \cite{DElia:1307.8063,Levkova:1309.1142,Braguta:1909.09547,Bali:1406.0269,Endrodi:2004.08778} and analytic predictions \cite{Cassing:1312.3189,Kanazawa:1410.6253,BallonBayona:2005.00500} we found paramagnetic behavior at large temperatures $T > T_c$. At low temperatures the $SU\lr{2}$ gauge theory also appears to be paramagnetic, although for the second-lowest temperature ($L_t = 20$) the magnetic susceptibility is zero within statistical error. We cannot therefore exclude the weak diamagnetism scenario \cite{Endrodi:2004.08778,Hofmann:2103.04937} at low temperatures. As stressed in \cite{Endrodi:2004.08778}, careful extrapolation to the continuum limit is required to obtain the diamagnetic response, which we leave for future work. With our discrete set of $\mu$ values we might also miss the diamagnetic regime immediately above the diquark condensation threshold. This regime was found in $SU\lr{3}$ gauge theory at finite isospin chemical potential \cite{Endrodi:1407.1216}, and might also exist in $SU\lr{2}$ gauge theory because the conventional chemical potential and isospin chemical potential are equivalent for $SU\lr{2}$ gauge group \cite{Kogut:hep-ph/0001171}.

More simulations are required to study these regimes. At higher temperatures our results for $\chi\lr{T, \mu}$ are close to the lattice QCD results \cite{Bali:1406.0269,Endrodi:2004.08778}.

We find that at all temperatures finite chemical potential tends to make the paramagnetic response stronger. Our estimates for the first coefficient of the expansion of the magnetic susceptibility $\chi\lr{T, \mu}$ in even powers of $\mu/T$ are close to the free fermion results and lie in the range $3 \cdot 10^{-3} \ldots 5 \cdot 10^{-3}$.

The paramagnetic response turns out to be particularly strong at $\mu > m_{\pi}/2$, and is practically temperature-independent in the deconfined regime. As we enter the diquark condensation phase (the lowest temperature on Fig.~\ref{fig:chi}), the magnetic susceptibility significantly increases. This suggests that the diquark condensation phase in $SU\lr{2}$ gauge theory might exhibit quite strong paramagnetism at least in some range of $T$ and $\mu$ values.

\begin{acknowledgements}
D.~S. and L.~v.~S. were supported by the Helmholtz International Center (HIC) for FAIR. D.~S. also received funding from the European Union's Horizon 2020 research and innovation programme under grant agreement No.~871072, also known as CREMLINplus (Connecting Russian and European Measures for Large-scale Research Infrastructures).

This work was performed using the Cambridge Service for Data Driven Discovery (CSD3), part of which is operated by the University of Cambridge Research Computing on behalf of the STFC DiRAC HPC Facility (www.dirac.ac.uk). The DiRAC component of CSD3 was funded by BEIS capital funding via STFC capital grants ST/P002307/1 and ST/R002452/1 and STFC operations grant ST/R00689X/1. DiRAC is part of the National e-Infrastructure.

The simulations were also performed on the GPU cluster at the Institute for Theoretical Physics at Giessen University.
\end{acknowledgements}


\begin{thebibliography}{10}
\providecommand{\url}[1]{{#1}}
\providecommand{\urlprefix}{URL }
\expandafter\ifx\csname urlstyle\endcsname\relax
  \providecommand{\doi}[1]{DOI \discretionary{}{}{}#1}\else
  \providecommand{\doi}{DOI \discretionary{}{}{}\begingroup
  \urlstyle{rm}\Url}\fi

\bibitem{BlandfordMagneticSusceptibility}
R.D. Blandford, L.~Hernquist, J.~Phys.~C:~{Solid State Phys.} \textbf{15}, 6233
   (1982).
\newblock \urlprefix\url{http://dx.doi.org/10.1088/0022-3719/15/30/017}

\bibitem{Endrodi:1407.1216}
G.~Endr\H{o}di, Phys.~Rev.~D \textbf{90}, 094501 (2014).
\newblock \urlprefix\url{https://dx.doi.org/10.1103/PhysRevD.90.094501},
  \urlprefix\url{https://arxiv.org/abs/1407.1216}

\bibitem{Bali:1311.2559}
G.S. Bali, F.~Bruckmann, G.~Endr\H{o}di, A.~Sch{\"{a}}fer, Phys.~Rev.~Lett.
  \textbf{112}, 042301 (2014).
\newblock \doi{https://dx.doi.org/10.1103/PhysRevLett.112.042301}.
\newblock \urlprefix\url{https://arxiv.org/abs/1311.2559}

\bibitem{DElia:1307.8063}
C.~Bonati, M.~{D'Elia}, M.~Mariti, F.~Negro, F.~Sanfilippo, Phys.~Rev.~Lett.
  \textbf{111}, 182001 (2013).
\newblock \urlprefix\url{http://dx.doi.org/10.1103/PhysRevLett.111.182001},
  \urlprefix\url{https://arxiv.org/abs/1307.8063}

\bibitem{Levkova:1309.1142}
L.~Levkova, C.~DeTar, Phys.~Rev.~Lett. \textbf{112,}, 012002 (2014).
\newblock \urlprefix\url{http://dx.doi.org/10.1103/PhysRevLett.112.012002},
  \urlprefix\url{https://arxiv.org/abs/1309.1142}

\bibitem{Braguta:1909.09547}
V.V. Braguta, M.N. Chernodub, A.Y. Kotov, A.V. Molochkov, A.A. Nikolaev,
  Phys.~Rev.~D \textbf{100}, 114503 (2019).
\newblock \urlprefix\url{http://dx.doi.org/10.1103/PhysRevD.100.114503},
  \urlprefix\url{https://arxiv.org/abs/1909.09547}

\bibitem{Endrodi:1301.1307}
G.~Endrodi, JHEP \textbf{1304}, 023 (2013).
\newblock \urlprefix\url{http://dx.doi.org/10.1007/JHEP04(2013)023},
  \urlprefix\url{https://arxiv.org/abs/1301.1307}

\bibitem{Bali:1406.0269}
G.S. Bali, F.~Bruckmann, G.~Endrodi, S.D. Katz, A.~Schafer, JHEP \textbf{1408},
  177 (2014).
\newblock \urlprefix\url{https://dx.doi.org/10.1007/JHEP08(2014)177},
  \urlprefix\url{https://arxiv.org/abs/1406.0269}

\bibitem{Endrodi:2004.08778}
G.S. Bali, G.~Endrődi, S.~Piemonte, JHEP \textbf{07}, 183 (2020).
\newblock \urlprefix\url{http://dx.doi.org/10.1007/JHEP07(2020)183},
  \urlprefix\url{https://arxiv.org/abs/2004.08778}

\bibitem{Hofmann:2103.04937}
C.P. Hofmann.
\newblock Diamagnetic and paramagnetic phases in low-energy quantum
  chromodynamics (2021).
\newblock \urlprefix\url{https://arxiv.org/abs/2103.04937}

\bibitem{Cassing:1312.3189}
T.~Steinert, W.~Cassing, Phys.~Rev.~C \textbf{89}, 035203 (2014).
\newblock \urlprefix\url{http://dx.doi.org/10.1103/PhysRevC.89.035203},
  \urlprefix\url{https://arxiv.org/abs/1312.3189}

\bibitem{Kanazawa:1410.6253}
K.~Kamikado, T.~Kanazawa, JHEP \textbf{1501}, 129 (2015).
\newblock \urlprefix\url{http://dx.doi.org/10.1007/JHEP01(2015)129},
  \urlprefix\url{https://arxiv.org/abs/1410.6253}

\bibitem{BallonBayona:2005.00500}
A.~{Ballon-Bayona}, J.P. Shock, D.~Zoakos, JHEP \textbf{2010}, 193 (2020).
\newblock \urlprefix\url{http://dx.doi.org/10.1007/JHEP10(2020)193},
  \urlprefix\url{https://arxiv.org/abs/2005.00500}

\bibitem{Ghosh:2103.08407}
R.~Ghosh, B.~Karmakar, M.~{Golam~Mustafa}.
\newblock Chiral susceptibility in dense thermo-magnetic {QCD} medium within
  {HTL} approximation (2021).
\newblock \urlprefix\url{https://arxiv.org/abs/2103.08407}

\bibitem{Bergman:0806.0366}
O.~Bergman, G.~Lifschytz, M.~Lippert, Phys.~Rev.~D \textbf{79}, 105024 (2009).
\newblock \urlprefix\url{http://dx.doi.org/10.1103/PhysRevD.79.105024},
  \urlprefix\url{https://arxiv.org/abs/0806.0366}

\bibitem{Tatsumi:1008.3753}
T.~Tatsumi, J.~Phys.~Conf.~Ser. \textbf{312}, 012014 (2011).
\newblock \urlprefix\url{http://dx.doi.org/10.1088/1742-6596/312/1/012014},
  \urlprefix\url{https://arxiv.org/abs/1008.3753}

\bibitem{Kogut:hep-lat/0105026}
J.B. Kogut, D.K. Sinclair, S.J. Hands, S.E. Morrison, Phys.~Rev.~D \textbf{64},
  094505 (2001).
\newblock \urlprefix\url{http://dx.doi.org/10.1103/PhysRevD.64.094505},
  \urlprefix\url{https://arxiv.org/abs/hep-lat/0105026}

\bibitem{Kogut:hep-ph/0001171}
J.B. Kogut, M.A. Stephanov, D.~Toublan, J.J.M. Verbaarschot, A.~Zhitnitsky,
  Nucl.~Phys.~B \textbf{582}, 477  (2000).
\newblock \urlprefix\url{http://dx.doi.org/10.1016/S0550-3213(00)00242-X},
  \urlprefix\url{https://arxiv.org/abs/hep-ph/0001171}

\bibitem{Pisarski:0706.2191}
L.~McLerran, R.D. Pisarski, Nucl.~Phys.~A \textbf{796}, 83  (2007).
\newblock \urlprefix\url{http://dx.doi.org/10.1016/j.nuclphysa.2007.08.013},
  \urlprefix\url{https://arxiv.org/abs/0706.2191}

\bibitem{Braguta:1605.04090}
V.V. Braguta, E.~Ilgenfritz, A.Y. Kotov, A.V. Molochkov, A.A. Nikolaev,
  Phys.~Rev.~D \textbf{94}, 114510 (2016).
\newblock \urlprefix\url{http://dx.doi.org/10.1103/PhysRevD.94.114510},
  \urlprefix\url{https://arxiv.org/abs/1605.04090}

\bibitem{Son:hep-ph/0005225}
D.T. Son, M.A. Stephanov, Phys.~Rev.~Lett. \textbf{86}, 592 (2001).
\newblock \urlprefix\url{https://dx.doi.org/10.1103/PhysRevLett.86.592},
  \urlprefix\url{https://arxiv.org/abs/hep-ph/0005225}

\bibitem{GiulianiVignaleElectronLiquid}
G.~Giuliani, G.~Vignale, \emph{Quantum Theory of the Electron Liquid}
  (Cambridge University Press, 2012).
\newblock \urlprefix\url{https://dx.doi.org/10.1017/CBO9780511619915}

\bibitem{Buividovich:20:1}
P.V. Buividovich, L.~{von~Smekal}, D.~Smith, Phys.~Rev.~D \textbf{102}, 094510
  (2020).
\newblock \urlprefix\url{http://dx.doi.org/10.1103/PhysRevD.102.094510},
  \urlprefix\url{https://arxiv.org/abs/2007.05639}

\bibitem{Buividovich:20:2}
P.V. Buividovich, D.~Smith, L.~{von~Smekal}.
\newblock A numerical study of chiral separation effect in finite-density
  {SU(2)} gauge theory with dynamical fermions (2020).
\newblock \urlprefix\url{https://arxiv.org/abs/2012.05184}

\bibitem{Hasenfratz:hep-lat/0103029}
A.~Hasenfratz, F.~Knechtli, Phys.~Rev.~D \textbf{64}, 034504 (2001).
\newblock \urlprefix\url{http://dx.doi.org/10.1103/PhysRevD.64.034504},
  \urlprefix\url{https://arxiv.org/abs/hep-lat/0103029}

\bibitem{Nikolaev:2008.12326}
G.~Aarts, A.~Nikolaev.
\newblock Electrical conductivity of the quark-gluon plasma: perspective from
  lattice {QCD} (2020).
\newblock \urlprefix\url{https://arxiv.org/abs/2008.12326}

\end{thebibliography}

\end{document}